\documentclass[aps,prd,onecolumn,showpacs,groupedaddress,nofootinbib]{revtex4-2}
\usepackage{graphicx}
\usepackage{dcolumn}
\usepackage{bm}
\usepackage{color}
\usepackage{longtable}
\usepackage{supertabular}
\usepackage[T1]{fontenc}
\usepackage{epstopdf}
\usepackage{amsmath}
\usepackage{amssymb}
\usepackage{setspace}
\usepackage{multirow}
\usepackage{booktabs}
\usepackage[section]{placeins}
\usepackage{mathrsfs}
\usepackage{hyperref}

\makeatletter

\newcommand{\Rmnum}[1]{\expandafter\@slowromancap\romannumeral #1@} 
\newcommand{\bq}{\begin{equation}}
\newcommand{\eq}{\end{equation}}
\newcommand{\bqn}{\begin{eqnarray}}
\newcommand{\eqn}{\end{eqnarray}}
\newcommand{\nb}{\nonumber}

\makeatother

\begin{document}
\title{On partition temperature of massless particles in high-energy collisions}

\author{Wei-Liang Qian\textsuperscript{2,1,3}}\email[E-mail: ]{wlqian@usp.br}
\author{Kai Lin\textsuperscript{4,1}}
\author{Rui-Hong Yue\textsuperscript{1}}
\author{Yogiro Hama\textsuperscript{5}}\email[E-mail: ]{hama1936@gmail.com}
\author{Takeshi Kodama\textsuperscript{6,7}}

\affiliation{$^{1}$ Center for Gravitation and Cosmology, College of Physical Science and Technology, Yangzhou University, Yangzhou 225009, China}
\affiliation{$^{2}$ Escola de Engenharia de Lorena, Universidade de S\~ao Paulo, 12602-810, Lorena, SP, Brazil}
\affiliation{$^{3}$ Faculdade de Engenharia de Guaratinguet\'a, Universidade Estadual Paulista, 12516-410, Guaratinguet\'a, SP, Brazil}
\affiliation{$^{4}$ Hubei Subsurface Multi-scale Imaging Key Laboratory, School of Geophysics and Geomatics, China University of Geosciences, 430074, Wuhan, Hubei, China}
\affiliation{$^{5}$ Instituto de F\'isica, Universidade de S\~ao Paulo, C.P. 66318, 05315-970, S\~ao Paulo-SP, Brazil}
\affiliation{$^{6}$ Instituto de F\'isica, Universidade Federal do Rio de Janeiro, C.P. 68528, 21945-970, Rio de Janeiro-RJ , Brazil}
\affiliation{$^{7}$ Instituto de F\'isica, Universidade Federal Fluminense, 24210-346, Niter\'oi-RJ, Brazil}

\begin{abstract}
Although partition temperature derived using the Darwin-Fowler method is exact for simple scenarios, the derivation for complex systems might reside on specific approximations whose viability is not ensured if the thermodynamic limit is not attained. 
This work elaborates on a related problem relevant to relativistic high-energy collisions.
On the one hand, it is simple enough that closed form expression can be obtained precisely for the one-particle distribution function.
On the other hand, the resulting expression is not an exponential form, and therefore it is not straightforward that the notion of partition function could be implied.
Specifically, we derive the one-particle distribution function for massless particles where the phase space integration is performed exactly for the underlying canonical ensemble consisting of a given number of particles.
We discuss the viability of the partition temperature in this case.
Possible implications of the obtained results regarding the observed Tsallis distribution in transverse momentum spectra in high-energy collisions are also addressed.
\end{abstract}

\date{July 31st, 2023}

\maketitle

\newpage
\section{Introduction}\label{sec1}

In high-energy hadronic collisions, the outgoing particle that possesses the same quantum numbers as the incoming one is known as the leading particle, whose identification allows us to obtain additional details about the scattering process kinematics.
For high-energy collisions~\cite{RHIC-Lee-01, RHIC-Lee-02, RHIC-Lee-review-03}, it was proposed that for the semi-inclusive scattering processes~\cite{SPS-review-05, JLAB-review-05, EIC-review-04, HERA-review-05}, the single-particle distribution for (all of the remaining) nonleading particles can be derived from the joint exclusive probability distribution~\cite{jet-ph-05}.
Specifically, the latter is essentially a micro-canonical ensemble consisting of $n$ particles under the relevant conservation laws, which reads
\begin{eqnarray}
\mathcal{P} =\prod_{l=1}^n \frac{d^3\mathbf{p}_l}{E_l}g(\mathbf{p}_1,\mathbf{p}_2,\cdots,\mathbf{p}_n)\delta\left[\sum_{i=1}^n p_{Li} - P_L\right] \delta^{2}\left[\sum_{j=1}^n \mathbf{p}_{Tj} - \mathbf{P}_T\right] \delta\left[\sum_{k=1}^n E_k - W\right] ,\label{PmeFormal}
\end{eqnarray}
where $W$, $P_L$, and $\mathbf{P}_T$, are the total energy, longitudinal and transversal momenta of the $N$-particle system in the center-of-mass frame, 
$g(\mathbf{p}_1,\cdots,\mathbf{p}_n)$ refers to some additional constraint of the phase space, whose form in this study will be made explicit below in Eq.~\eqref{fSForm}.
To focus on the single-particle spectrum, one integrates out the degrees of freedom associated with the remaining $n-1$ particles to find
\begin{eqnarray}
\frac{d^3\bar{n}}{dy d\mathbf{p}_T} = \int \prod_{l=2}^n \frac{d^3\mathbf{p}_l}{E_l}g(\mathbf{p},\mathbf{p}_2,\cdots,\mathbf{p}_n)\delta\left[\sum_{i=2}^n p_{Li} +p_{L} - P_L\right] \delta^{2}\left[\sum_{j=2}^n \mathbf{p}_{Tj}+\mathbf{p}_{T} - \mathbf{P}_T\right] \delta\left[\sum_{k=2}^n E_k + E - W\right] ,\label{n1Formal}
\end{eqnarray}
one manifestly arrives at a canonical ensemble
\begin{eqnarray}
\frac{d^3\bar{n}}{dy d\mathbf{p}_T} \propto \exp\left(-\frac{E}{T_p}\right) ,\label{nTpGenForm}
\end{eqnarray}
for a single-particle with longitudinal and transverse momenta $p_L$ and $\mathbf{p}_T$, rapidity $y=\ln\frac{E+p_L}{E-p_L}$, and energy $E$,
where $T_p$ effectively plays the role of temperature and is referred to as partition temperature~\cite{book-statistical-mechanics-huang}.
Such an endeavor was explicitly carried out in~\cite{jet-ph-05}~\footnote{The calculations performed in~\cite{jet-ph-05}, which leads to Eq.~\eqref{nTpGenForm}, primarily employed the following assumptions and approximations.
(1) The momentum conservation in the transverse direction is largely ignored.
(2) A specific form of the density function $f(y, \mathbf{p}_T)$ is assumed, c.f. Eq.~\eqref{n1Spec}.
(3) It was argued that the integral in $\eta=\sqrt{(\beta+s)^2-t^2}$ (where some relevant variables are defined in Eqs.~\eqref{deltaEpL}) is dominated by the region $\eta\to 0$, and thus an approximate form is used.
(4) The saddle approximation is employed for the integral in $\zeta=\mathrm{arctanh}\left(\frac{t}{\beta+s}\right)$.}, 
and the obtained analytic results on the partition temperatures were compared against those obtained by fitting to the experimental data for $p\bar{p}$ collisions at 540 GeV~\cite{jet-ph-02}.
By adopting the parameters extracted from the data, excellent agreement was achieved, where, for the most part, the discrepancies are less than 5\%.

In principle, the Boltzmann-Gibbs characteristic of the above results can be understood~\cite{jet-ph-02, jet-ph-03} in terms of Darwin-Fowler framework~\cite{book-statistical-mechanics-huang}.
In this regard, the notion of partition temperature $T_p$ emerges from a contour integration in terms of a {\it selector variable}~\cite{statistics-Darwin-Fowler-09}, which is utilized to enforce the relevant conservation laws.
Specifically, the common parameter between different assemblies resulting from such an approach bears a resemblance to the temperature and is referred to in the literature as partition temperature.
It is noted that the feasibility of the Darwin-Fowler formalism largely resides in the validity of the saddle point method, which is often employed to approximate the contour integration.
In particular, the partition function is essentially governed by the location of the saddle point.
Even thought the partition is obtained without referring to the thermodynamic equilibrum, its link to other thermodynamic quantities such as entropy has been elaborated~\cite{statistics-Darwin-Fowler-02}.
In this approach, the saddle point corresponds to a strong maximum on the contour in question, which is true when the thermodynamic limit is attained.
However, if the relevant degrees of freedom are not significant, numerical calculations indicated that further caution might be taken~\cite{statistics-Darwin-Fowler-25}. 
To be more precise, the derivation based on the approximation and, subsequently, the notion of partition temperature might cease to be valid. 
Calculations need to be performed to validate the partition temperature for the specific scenario in question.
As discussed below, the feasibility of partition temperature plays an essential role in the observed Tsallis distribution; therefore, the resultant particle distribution at small multiplicities is physically pertinent.

The calculations~\cite{jet-ph-05} carried out by explicit phase space integration also employed the saddle point method.
It was found that the divergence becomes more significant for smaller multiplicities.
When scrutinizing the results more closely, it was pointed out that~\cite{jet-ph-05} the results did not correctly reproduce the dependence of average transverse momentum on the event multiplicity, despite the apparent success of reproducing the partition temperature.
Moreover, the deviation increases further if one redefines the total energy using the experimental data on transverse momentum.
This, in turn, leads to some degree of speculation about whether the notion of partition temperature is indeed valid in such a context, such as $\bar{p}p$ collisions, where the overall multiplicity is not significant.

For a wide range of $pp$, $p\bar{p}$, $AA$, and $e^+e^-$ collisions, the transverse-momentum spectrum for the intermediate and high momentum range ($p_T \gtrsim 3$ GeV) cannot be described by a Boltzmann-Gibbs distribution.
However, the resulting Tsallis distribution can still be understood in the framework of superstatistics. 
Specifically, since the system is not homogeneous, one may obtain the Tsallis distribution by assuming that the observed spectrum corresponds to an ensemble average of subsystems where either the temperature~\cite{jet-ph-10} or multiplicity~\cite{jet-ph-12, jet-ph-13} fluctuates.
It can be shown that an additional parameter, $q$, associated with the Tsallis-like distribution, measures the strength of fluctuations in either temperature or multiplicity.
In this context, the notion of partition function continues to play a significant role in these studies.

The present study is motivated by the above discussions.
We aim at a scenario where the contour integral can be performed precisely to steer clear of the uncertainty related to the saddle point approximation when the multiplicity is small. 
To this end, we elaborate on the case of massless particles where the phase space integration can be evaluated without saddle point approximation.
In this regard, the obtained result furnishes another explicit example where the concept of partition temperature is valid.

The remainder of the paper is organized as follows.
In the following section, we give a detailed account of the general form of the phase space integration that gives rise to the one-particle distribution from the microcanonical ensemble, essentially based on results obtained in Ref.~\cite{jet-ph-05}.
In Sec.~\ref{sec3}, the specific contour integration is carried out precisely, and the results are presented.
We conduct numerical analysis in Sec.~\ref{sec4} and discuss the viability of the extracted partition temperatures.
Further discussions and concluding remarks are given in the last section.

\section{Phase-space integration of the one-particle momentum distribution}\label{sec2}

We start by considering a system of $N$ particles with total energy $W$, momentum $\mathbf{P}$, and subsequently total invariant mass $M=\sqrt{W^2-\mathbf{P}^2}$.
The particle's mass is denoted by $m$, which will be taken to be zero.

Following~\cite{jet-ph-05}, we divide the phase space into $N$ small intervals whose occupation numbers will be denoted as $n_\ell$ with $\ell=1,2,\cdots,N$.
The probability of finding a microscopic distribution $\{n_\ell\}\equiv \{n_1,\cdots,n_N\}$, Eq.~\eqref{PmeFormal} takes the form
\begin{eqnarray}
\mathcal{P}\left(\{n_\ell\}\right) =\frac{n!}{n_1!\cdots n_N!}q_1^{n_1}\cdots q_N^{n_N} \delta_{\left(n,\sum_{\ell=1}^N n_\ell\right)}\delta\left[\sum_{i=1}^N n_ip_{Li} - P_L\right] \delta^2\left[\sum_{j=1}^N n_j\mathbf{p}_{Tj} - \mathbf{P}_T\right] \delta\left[\sum_{k=1}^N n_k E_k - W\right] ,\label{PmicroEn}
\end{eqnarray}
where $q_\ell$ stands for the probability of finding a particle in the phase space interval $\ell$.
Apparently, $q_\ell$ is proportional to the volume of the phase space $dV=dyd\mathbf{p}_{T}$, so that $q_\ell=f(y_\ell,\mathbf{p}_{T\ell})dyd\mathbf{p}_{T}$, where the probability density function $f(y,\mathbf{p}_{T})$ is normalized
\begin{eqnarray}
\sum_{\ell=1}^N q_\ell=1 \to \lim\limits_{N\to\infty}\sum_{\ell=1}^N q_\ell= \int f(y,\mathbf{p}_{T})dyd\mathbf{p}_{T} = 1 .\label{fNorm}
\end{eqnarray}
One takes $f(y,\mathbf{p}_{T})=1$, if the occupation probability is simply proportional to the volume of the phase space.

Using the probability given by Eq.~\eqref{PmicroEn}, the single particle momentum Eq.~\eqref{n1Formal} distribution reads
\begin{eqnarray}
\left.\frac{d^3\bar{n}}{dyd\mathbf{p}_T}\right|_{n,W,P}
\equiv\langle n_k \rangle 
=\frac{\sum_{\{n_\ell\}}P(\{n_\ell\})n_k}{\sum_{\{n_\ell\}}P(\{n_\ell\})}\equiv \frac{A}{B} .\label{n1Spec}
\end{eqnarray}
We note that both Eqs.~\eqref{PmicroEn} and~\eqref{n1Spec} have explicitly taken into account the conservation of the total number of particles, momentum, and energy.

To deal with the conservation in energy and longitudinal momentum, one introduces two Laplace transforms in the variables $s$ and $t$ and rewrites these quantities in terms of transverse momentum and rapidity, namely, $E=\sqrt{\mathbf{p}_T^2+m^2}\cosh y$ and $p_L=\sqrt{\mathbf{p}_T^2+m^2}\sinh y$. 
Specifically, We have
\begin{eqnarray}
\delta\left[\sum_{k=1}^N n_kE_k - W\right] &=& \int_{\epsilon_0-i\infty}^{\epsilon_0+i\infty} ds \exp\left[\left(W-\sum_{k=1}^N n_k E_k\right)s\right]  
=\int_{\epsilon_0-i\infty}^{\epsilon_0+i\infty} ds e^{Ws}\prod_{k=1}^N \left[e^{-  E_k s}\right]^{n_k} \nb\\
\delta\left[\sum_{i=1}^N n_ip_{Li} - P_L\right] &=& \int_{\epsilon_1-i\infty}^{\epsilon_1+i\infty} dt \exp\left[-\left(P_L-\sum_{i=1}^N n_i p_{Ti}\right)t\right] 
=\int_{\epsilon_1-i\infty}^{\epsilon_1+i\infty} dt e^{-P_L t}\prod_{i=1}^N \left[e^{  p_{Li} t}\right]^{n_i} ,\label{deltaEpL}
\end{eqnarray}
where $\epsilon_0 = \Re s > \Re t = \epsilon_1\ge 0$, so that the integrals are convergent.
The conservation law for the transversal momentum can be treated similarly by introducing the Fourier transform of a two-dimensional delta function
\begin{eqnarray}
\delta^{2}\left[\sum_{j=1}^n \mathbf{p}_{Tj} - \mathbf{P}_T\right]
=\int d\mathbf{u}_T \exp\left[-i\left(\mathbf{P}_T-\sum_{j=1}^N n_j \mathbf{p}_{Tj}\right)\cdot \mathbf{u}_T\right] 
=\int d\mathbf{u}_T e^{- i\mathbf{P}_T \cdot \mathbf{u}_T}\prod_{j=1}^N \left[e^{i \mathbf{p}_{Tj}\cdot \mathbf{u}_T}\right]^{n_j}  .\label{deltapT}
\end{eqnarray}

The Kronecker delta responsible for particle number conservation can be dealt with by noticing the relation
\begin{eqnarray}
\delta_{\left(n,\sum_{\ell=1}^N n_\ell\right)} 
= \int_0^{2\pi} dv \exp\left[-i\left(n-\sum_{\ell=1}^N n_\ell\right)v\right] .\label{deltaN}
\end{eqnarray}

By plugging Eqs.~\eqref{deltaEpL}, \eqref{deltapT}, and~\eqref{deltaN} into the numerator of Eq.~\eqref{n1Spec} and picking out the terms related to the occupation numbers, we have
\begin{eqnarray}
\left.\frac{d^3\bar{n}}{dyd\mathbf{p}_T}\right|_{n,W,P}
\equiv\langle n_k \rangle 
\simeq nf(y,\mathbf{p}_T)dyd\mathbf{p}_T\frac{C}{D} ,\label{nSim}
\end{eqnarray}
where
\begin{eqnarray}
C = \int_{\epsilon_0-i\infty}^{\epsilon_0+i\infty}ds\int_{\epsilon_1-i\infty}^{\epsilon_1+i\infty}dt\left[F(s,t)\right]^{n-1} 
\times \exp\left[(W-{p}_T\cosh y)s-(P_L-{p}_T\sinh y)t\right] ,\label{nCSimZero}
\end{eqnarray}
\begin{eqnarray}
D=\int_{\epsilon_0-i\infty}^{\epsilon_0+i\infty}ds\int_{\epsilon_1-i\infty}^{\epsilon_1+i\infty}dt\left[F(s,t)\right]^{n} \exp\left[Ws-P_Lt\right] ,\label{nDSimZero}
\end{eqnarray}
and
\begin{eqnarray}
F(s,t,\mathbf{u}_T) 
=dyd\mathbf{p}_Tf(y,\mathbf{p}_T) \exp\left[-{p}_T(s \cosh y- t \sinh y)+i\mathbf{p}_T \cdot \mathbf{u}_T\right]. \label{FdefZero} 
\end{eqnarray}
We relegate the details of the derivations of Eqs.~\eqref{nCSimZero} and~\eqref{nDSimZero} to App.~\ref{appA}.

It is noted that the integrations in $s$ and $t$ given by Eqs.~\eqref{nCSim},~\eqref{nDSim},~\eqref{nCSimZero}, and~\eqref{nDSimZero} still pose a challenge for most scenarios.
For most cases, the approximation method is indispensable.
In particular, saddle point approximation was employed for the ansatz in~\cite{jet-ph-05}.

\section{The contour integration}\label{sec3}

In this section, we elaborate on a simplified scenario that remains physically relevant while avoiding the need for any approximations. 
Specifically, we constrain the emission of massless particles entirely to the longitudinal direction, with the probability density function taking the form
\begin{eqnarray}
f(p)=\alpha e^{-\beta|p|},\label{fSForm}
\end{eqnarray}
where $\beta>0$ and $\alpha=\beta/2$ serves as the normalization constant.
Notably, the parameter $\beta$ should not be construed as a ``temperature''.
As it turns out, the specific form of Eq.~\eqref{fSForm} simplifies the integrations, and its role is primarily technical to ensure the convergence of the integrations. 
Furthermore, as discussed below, the {\it pure} phase space integral can be readily recovered by taking the limit $\beta\to 0_+$ at the end of the calculations.

Subsequently, $F(s,t)$ defined by Eq.~\eqref{Fdef} possesses the following form
\begin{eqnarray}
F(s, t)&=&\frac{\beta}{2}\int_{-\infty}^{\infty} \exp\left[-(\beta+s)|p|+pt\right], \nb\\
&=&\frac{\beta}{2}\left[\int_{-\infty}^{0} \exp\left[(t+\beta+s)p\right]+\int_{0}^{\infty} \exp\left[-(\beta+s-t)p\right]\right], \nb\\
&=&\frac{\beta}{2}\left[\frac{1}{t+\beta+s}+\frac{1}{t-\beta-s}\right], \nb\\
&=& \frac{\beta(\beta+s)}{(\beta+s)^2-t^2} ,\label{FstSimplified}
\end{eqnarray}
where one requires $\Re s > \Re t \ge 0$ so that the integrals are well-defined.
We note the simplification brought to Eq.~\eqref{Fdef} in this case, which also benefits from the form given by Eq.~\eqref{fSForm}.
On the contrary, the exponential cannot be simplified for massive particles, which eventually leads to further change of variables and the saddle approximation.

Now we proceed to evaluate $C$ and $D$ defined by Eqs.~\eqref{nCSim} and~\eqref{nDSim}.
For the denominator, we have 
\begin{eqnarray}
D &=& \int_{\epsilon_0-i\infty}^{\epsilon_0+i\infty}ds\int_{\epsilon_1-i\infty}^{\epsilon_1+i\infty}dt\left[\frac{\beta(\beta+s)}{(\beta+s)^2-t^2}\right]^n\exp(Ms) \nb\\
&=&\beta^n \int_{\epsilon_0-i\infty}^{\epsilon_0+i\infty}ds (\beta+s)^n \exp(Ms)\int_{\epsilon_1-i\infty}^{\epsilon_1+i\infty}dt G(t) .\nb\\
\end{eqnarray}
where
\begin{eqnarray}
G(t)=\left[\frac{1}{(\beta+s)^2-t^2}\right]^n ,
\end{eqnarray}
and one has utilized the relation $W=M$ and $P=0$ in the center of mass frame of the jet.

The integral in $t$ can be done using the residue theorem by completing the contour in the counterclockwise direction on an infinite semi-circle.
The analytic function $G(t)$ has two $n$-th order poles.
By picking up the relevant one on the l.h.s. of the imaginary axis $t_0=-(\beta+s)$, one finds
\begin{eqnarray}
D &=&\beta^n \int_{\epsilon_0-i\infty}^{\epsilon_0+i\infty}ds (\beta+s)^n \exp(Ms)\int_{\epsilon_1-i\infty}^{\epsilon_1+i\infty}dt G(t) \nb\\
&=&\beta^n \int_{\epsilon_0-i\infty}^{\epsilon_0+i\infty}ds (\beta+s)^n \exp(Ms)\oint_C dt G(t) \nb\\
&=&\beta^n \int_{\epsilon_0-i\infty}^{\epsilon_0+i\infty}ds (\beta+s)^n \exp(Ms)2\pi i\mathrm{Res} (G, t_0) \nb\\
&=&\frac{2\pi i\beta^n (2n-2)!}{2^{2n-1}[(n-1)!]^2} \int_{\epsilon_0-i\infty}^{\epsilon_0+i\infty}ds H(s) ,\label{DintS}
\end{eqnarray}
where 
\begin{eqnarray}
H(s) = \frac{e^{Ms}}{(s+\beta)^{n-1}} ,
\end{eqnarray}
as it is straightforward to find
\begin{eqnarray}
\mathrm{Res} (G, t_0) &=& \frac{1}{(n-1)!} \lim_{t \to t_0} \frac{d^{n-1}}{dt^{n-1}} \left( (t-t_0)^n G(t) \right) \nb\\
&=& \frac{1}{(n-1)!} \lim_{t \to t_0} \frac{d^{n-1}}{dt^{n-1}} \frac{(-1)^n}{(t+t_0)^n} \nb\\
&=& \frac{(-1)^n}{(n-1)!} \frac{(-n)(-n-1)\cdots(-2n+2)}{(2t_0)^{2n-1}} \nb\\
&=& \frac{(2n-2)!}{[(n-1)!]^2[2(\beta+s)]^{2n-1}} .
\end{eqnarray}

The last integral of Eq.~\eqref{DintS} can again be performed using residue theorem by using Jordan's lemma and enclosing the contour in the counterclockwise direction on an infinite semi-circle.
By picking out the contribution from residual on the negative real axis $s=-\beta$, we have
\begin{eqnarray}
\oint_C H(s)ds &=& 2\pi i\mathrm{Res} (H, -\beta) \nb\\
&=& \frac{2\pi i}{(n-2)!} \lim_{s \to -\beta} \frac{d^{n-2}}{ds^{n-2}} \left( (s+\beta)^{n-2} H(s) \right) \nb\\
&=& \frac{2\pi i}{(n-2)!} \lim_{s \to -\beta} \frac{d^{n-2}}{ds^{n-2}} e^{Ms} \nb\\
&=& \frac{2\pi i M^{n-2}}{(n-2)!} e^{-\beta M}.
\end{eqnarray}
Therefore, one obtains
\begin{eqnarray}
D=\frac{-\pi^2(2n-2)!\beta^2(\beta M)^{n-2}}{2^{2n-3}[(n-1)!]^2(n-2)!}e^{-\beta M} .
\end{eqnarray}

The calculation of Eq.~\eqref{nCSim} can be carried out in a similar fashion, and we have
\begin{eqnarray}
C&=&\int_{\epsilon_0-i\infty}^{\epsilon_0+i\infty}ds\int_{\epsilon_1-i\infty}^{\epsilon_1+i\infty}dt\left[\frac{\beta(\beta+s)}{(\beta+s)^2-t^2}\right]^{n-1}\exp\left[(M-|p|)s+pt\right]\nb\\
&=&\beta^{n-1}\int_{\epsilon_0-i\infty}^{\epsilon_0+i\infty}ds (\beta+s)^{n-1}\exp\left[(M-|p|)s\right]
\int_{\epsilon_1-i\infty}^{\epsilon_1+i\infty}dt\frac{e^{pt}}{\left[(\beta+s)^2-t^2\right]^{n-1}}\nb\\
&=& 2\pi i\beta^{n-1}\sum_{r=0}^{n-2} \frac{(n+r-2)!}{r!(n-r-2)!}\frac{p^{n-2-r}e^{-p \beta}}{2^{n+r-1}} 
\int_{\epsilon_0-i\infty}^{\epsilon_0+i\infty}ds \frac{\exp\left[(M-|p|-p)s\right]}{(\beta+s)^r}\nb\\
&=&(2\pi i)^2\beta^{n-1}\sum_{r=0}^{n-2} \frac{(n+r-2)!}{r!(n-r-2)!}\frac{p^{n-2-r}e^{-p \beta}}{2^{n+r-1}}\frac{ (M-|p|-p)^{r-1}}{(r-1)!}e^{-\beta (M-|p|-p)}\nb\\
&=&-4\pi^2\beta^{n-1}e^{-\beta (M-|p|)} \sum_{r=0}^{n-2} \frac{(n+r-2)!}{2^{n+r-1}r!(r-1)!(n-r-2)!}{p^{n-2-r}}(M-|p|-p)^{r-1} .
\end{eqnarray}

In the above calculations, the integral in $t$ can be carried out for 
\begin{eqnarray}
G=G(t)=\frac{(-1)^{n-1}e^{pt}}{[t-(\beta+s)]^{n-1}[t+(\beta+s)]^{n-1}} ,
\end{eqnarray}
using Jaordan's lemma.
Both residues at $t_{1,2}=\pm(\beta+s)$ are relevant, as one must choose the counterclockwise (clockwise) contour when $p>0$ ($p<0$).
Specifically, for $p>0$, we have
\begin{eqnarray}
\mathrm{Res}(G, t_2)&=&\frac{(-1)^{n-2}}{(n-2)!}\lim\limits_{t\to t_2}\left[\frac{d^{n-2}}{dt^{n-2}}\frac{e^{pt}}{(t-t_1)^{n-1}}\right] \nb\\
&=&\frac{(-1)^{n-2}}{(n-2)!}\lim\limits_{t\to t_2}\sum_{r=0}^{n-2} \begin{pmatrix}n-2\\r\end{pmatrix} \left[\frac{d^{n-2-r}}{dt^{n-2-r}}e^{pt}\right]\left[\frac{d^r}{dt^r}\frac{1}{(t-t_1)^{n-1}}\right] \nb\\
&=&\frac{(-1)^{n-2}}{(n-2)!}\lim\limits_{t\to t_2}\sum_{r=0}^{n-2} \begin{pmatrix}n-2\\r\end{pmatrix} \left[p^{n-2-r}e^{pt}\right]\frac{(-1)^r(n-1)n\cdots(n+r-2)}{(t-t_1)^{n+r-1}} \nb\\
&=&\sum_{r=0}^{n-2} \begin{pmatrix}n-2\\r\end{pmatrix} \frac{n\cdots(n+r-2)p^{n-2-r}e^{-p (\beta+s)}}{2^{n+r-1}(\beta+s)^{n+r-1}}\nb\\
&=&\sum_{r=0}^{n-2} \frac{(n+r-2)!}{r!(n-r-2)!}\frac{p^{n-2-r}e^{-p (\beta+s)}}{2^{n+r-1}(\beta+s)^{n+r-1}} .
\end{eqnarray}

The integral in $s$, the exponential $(M-|p|-p)>0$, dictates the use of Jordan's lemma and The residue of
\begin{eqnarray}
H(s)=\frac{e^{(M-|p|-p)s}}{(\beta+s)^r}
\end{eqnarray}
at $s=-\beta$ is found to be
\begin{eqnarray}
\mathrm{Res}(H, -\beta)
&=&\frac{1}{(r-1)!}\lim\limits_{s\to -\beta}\left[\frac{d^{r-1}}{ds^{r-1}}e^{(M-|p|-p)s}\right] \nb\\
&=&\frac{ (M-|p|-p)^{r-1}}{(r-1)!}e^{-\beta (M-|p|-p)} .
\end{eqnarray}

We note that when $p<0$, the integral in $t$ leads to a factor with exponential dependence of $e^{p(\beta+s)}$.
A similar expression is obtained after performing the integration in $s$.

Putting the pieces together, we have
\begin{eqnarray}
\left.\frac{d\bar{n}}{dp}\right|_{n,W,P}
\equiv\langle n_k \rangle 
= nf(p)dp\frac{C}{D} \propto \sum_{r=0}^{n-2}\frac{(n+r-2)!}{2^{n+r-1}r!(r-1)!(n-r-2)!}
\times p^{n-r-2}(M-|p|-p)^{r-1} , \label{dndpRes}
\end{eqnarray}
where it is noteworthy that the exponential term $\exp[-\beta|p|]$ cancels out identically.
In other words, the obtained one-particle distribution does not depend on the specific value of $\beta$, as taking the limit $\beta\to 0_+$ yields no changes to the resultant expression.
Additionally, based on the form of Eq.~\eqref{dndpRes}, it is unclear whether the resultant one-particle distribution can be characterized by any sort of partition temperature.

\section{The feasibility of partition temperature}\label{sec4}

The one-particle distribution function obtained in the last section, Eq.~\eqref{dndpRes}, possesses a rather tedious form consisting of a summation of polynomials in $p$.
In this section, we numerically evaluate the resulting distribution function and use a non-linear fit to extract the associated partition temperature.

\begin{figure}[ht]
\begin{tabular}{cc}
\begin{minipage}{250pt}
\centerline{\includegraphics[width=1.0\textwidth]{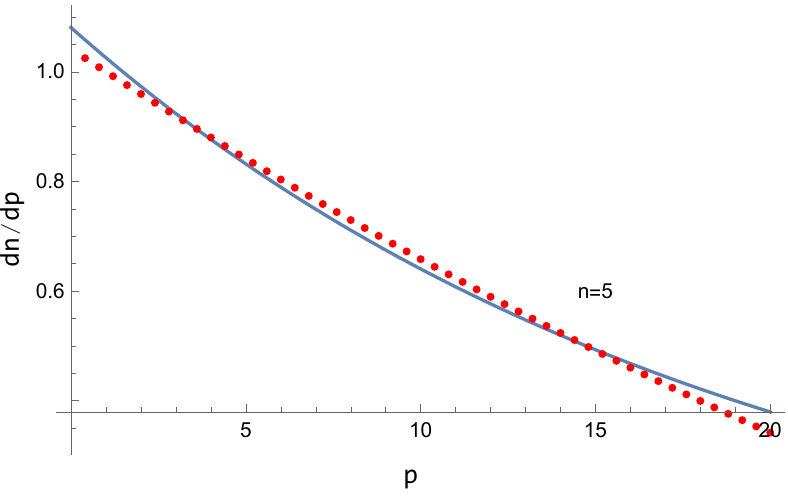}}
\end{minipage}
&
\begin{minipage}{250pt}
\centerline{\includegraphics[width=1.0\textwidth]{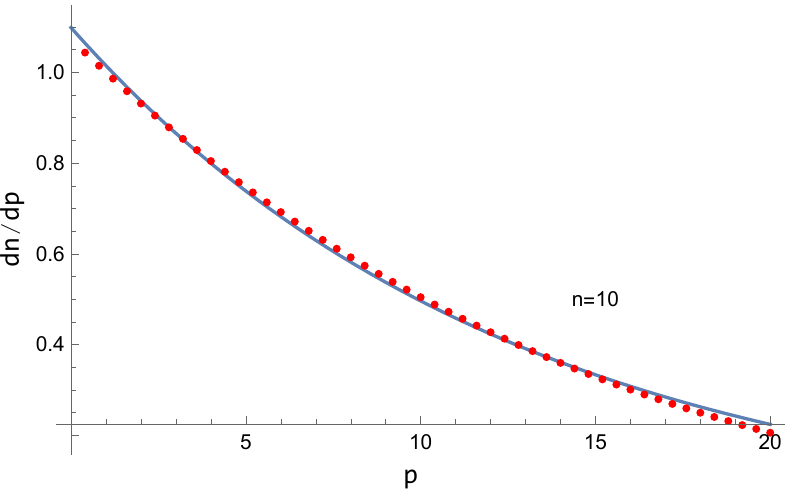}}
\end{minipage}\\
\begin{minipage}{250pt}
\centerline{\includegraphics[width=1.0\textwidth]{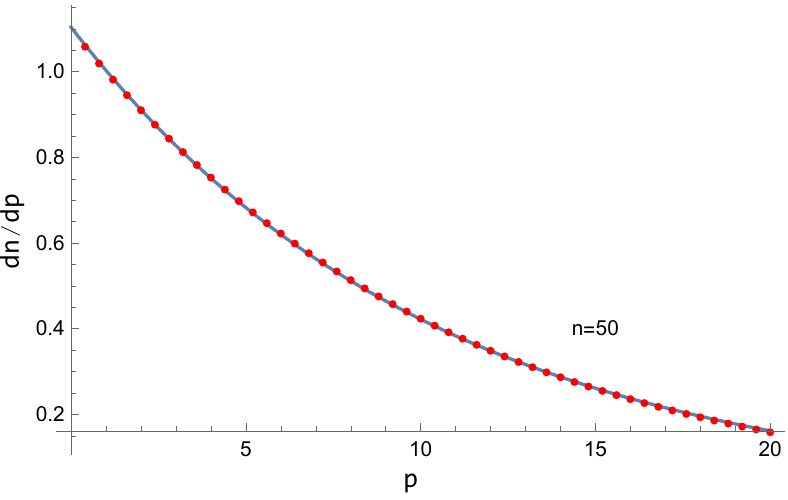}}
\end{minipage}
&
\begin{minipage}{250pt}
\centerline{\includegraphics[width=1.0\textwidth]{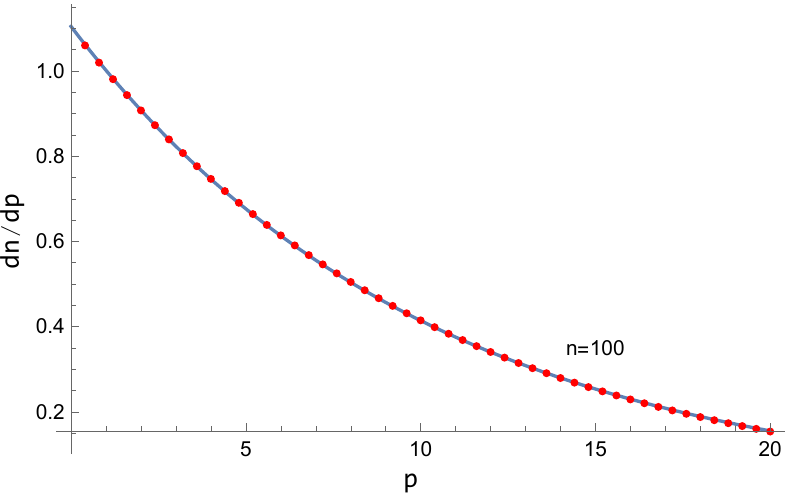}}
\end{minipage}
\end{tabular}
\renewcommand{\figurename}{Fig.}
\caption{The numerical fits (black curves) to the resulting one-particle distributions (solid red symbols) to the form $A\exp\left(- p/T_p\right)$ for different multiplicities are presented in the top-left ($n=5$), top-right ($n=10$), bottom-left ($n=50$), and bottom-right ($n=100$) plots, respectively.
For better visualization, the distribution functions are normalized by $\left.d\bar{n}/dp\right|_{p=1}$.}
\label{fig_nExpFit}
\end{figure}

\begin{figure}[ht]
\begin{tabular}{cc}
\begin{minipage}{250pt}
\centerline{\includegraphics[width=1.0\textwidth]{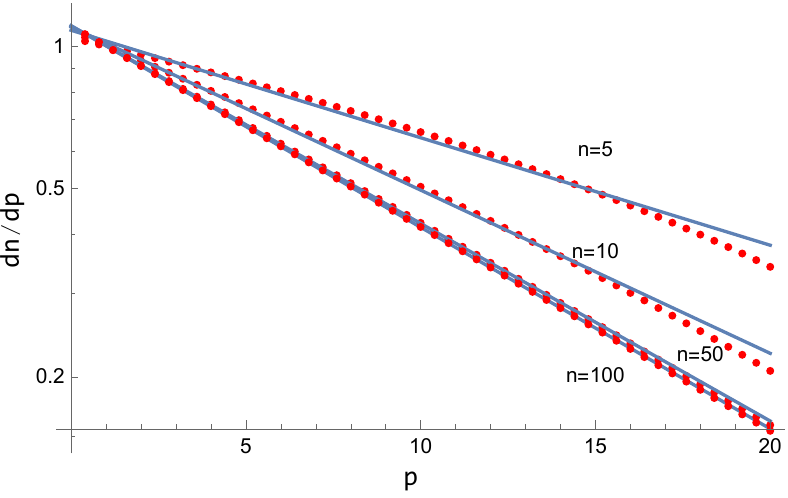}}
\end{minipage}
&
\begin{minipage}{250pt}
\centerline{\includegraphics[width=1.0\textwidth]{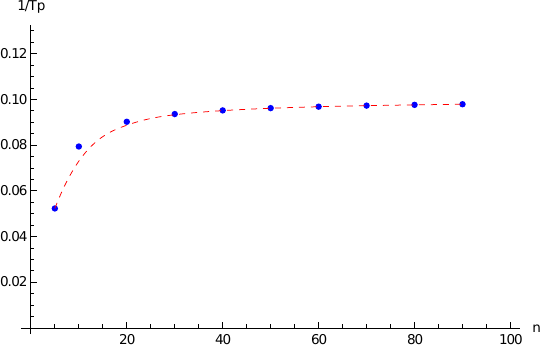}}
\end{minipage}
\end{tabular}{cc}
\renewcommand{\figurename}{Fig.}
\caption{
Left: The numerical fits shown in Fig.~\ref{fig_nExpFit} but presented in the logarithmic scale. 
From the top to bottom, the curves correspond to the multiplicities $n=5, 10, 50$, and $100$, respectively. 
Right: The extracted partition temperature $1/T_p$ as a function of the multiplicity $n$, where the red dashed curve is obtained by a third-order spline fit.}
\label{fig_TpVSn}
\end{figure}

The corresponding numerical results are shown in Figs.~\ref{fig_nExpFit} and~\ref{fig_TpVSn} and Tab.~\ref{tab_ExpFit}.
As the results mainly aim to demonstrate the feasibility of the effective temperature, the assumed values do not refer to any specific scenario, and natural units are used.
The calculations are carried out for a given average $\bar{p}$ so that the total mass read $M=W=n\bar{p}$.
The summations in Eq.~\eqref{dndpRes} are carried out numerically and sampled at $N_\mathrm{sample}=50$ points evenly distributed on the interval $p\in (0, 2\bar{p})$.
The calculations are for different multiplicities $n=5, 10, 50$, and $100$.
Also, without loss of generosity, we take $\bar{p}=10$.
We have confirmed that the qualitative conclusion drawn below remains unchanged, independent of this specific choice. 

In order to assess the validity of partition temperature, a nonlinear fit is carried out regarding the above range and depicted on top of the numerical results.
By using {\it Mathematica} build-in module $\mathrm{NonlinearModelFit}[]$, the resultant one-particle distribution function is fit to an exponential form $y=A\exp\left(- p/T_p\right)$.
The fit is performed under the assumption that the numerical values $y_i\ (i=1,\cdots,50)$ are independent and normally distributed with mean $\hat{y}_i$ and common standard deviation. 
The results are presented in Fig.~\ref{fig_nExpFit}, where black curves show the nonlinear fits and the numerical distribution functions are shown in solid red symbols.
It is somewhat surprising that even for $n=5$, the one-particle distribution function agrees reasonably well with the notion of partition temperature.
As the multiplicity increases, the agreement further improves, as shown in the left panel of Fig.~\ref{fig_TpVSn}.

More specific values of the extracted partition temperatures and the goodness of the fits are shown in Tab.~\ref{tab_ExpFit}.
As a measure of the fit's quality, the standard error is defined as the standard deviation divided by the square root of the number of samples.
Besides, the p-values are also evaluated, which give the probabilities, under the null hypothesis, of obtaining a result equal to or more extreme than the sampled points.
The smallness of both quantities indicates the partition temperature's feasibility; moreover, the fit becomes more reliable at large multiplicities.
For small multiplicity, however, the extracted value of partition temperature is not a constant.
This is understood as an indication that the quantity is irrelevant to thermodynamic equilibrium.
As the multiplicity increases, one observes that the partition temperature converges, as indicated by the right panel of Fig.~\ref{fig_TpVSn}.
It is intereting to point out that the effective temperature extracted numerically indicates that $T_p\to \bar{p}$ at the limit $n\to\infty$.
Nonetheless, it is also noted that the notion of partition function breaks down for the high-momentum region.

\begin{table*}[ht]
\caption{\label{tab_ExpFit} The extracted partition temperatures, the corresponding standard errors, and p-values for different multiplicities $n$.
The quality of the fit improves as multiplicity increases.}
\centering
\begin{tabular}{c cccccc}
\hline\hline
$n$                    &   5       &   10 &  30 & 40 & 50 & 100  \\
\hline\hline
${1}/{T_p}$        &~~~0.0523~~&~~~0.0794~~~&~~~0.0936~~~&~~~0.0952~~~&~~~0.0962~~~&~~~0.0981~~~\\
\hline
~~\text{standard error} & 0.0006   &  0.0005    &   0.0002   &  0.0001    &   0.0001   &   0.00005\\
\hline
~~\text{p-value}~~&~~ $1.2\times 10^{-51}$ ~~&~~ $9.5\times 10^{-67}$ ~~&~~ $3.0\times 10^{-91}$ ~~ &~~ $1.9\times 10^{-97}$~~ &~~ $3.2\times 10^{-102}$ ~~&~~   $6.5\times 10^{-117}$ \\
\hline\hline
\end{tabular}
\end{table*}

\section{Further discussions and concluding remarks}\label{sec5}

The notion of partition temperature born into the Darwin-Fowler approach is widely known, and the result is exact for simple scenarios.
As a result, it is often accepted and borrowed to different contexts without further scrutinization.
Nonetheless, for complex systems, the validity of the partition temperature resides on some approximations, such as the saddle point method, and subsequently, further caution should be taken.
The present study explores a scenario pertinent to particle emission in relativistic high-energy collisions, which possesses an analytic closed form for the resulting one-particle distribution.
It is observed that even the analytic result does not imply the existence of an effective temperature, and the distribution converges rapidly to an exponential form for particles with low and intermediate momenta.
Nonetheless, it is noted that the notion of partition temperature breaks down for the high-momentum region, due to the condition that the total energy is bounded.
We also note that the general scenario with massive particles and the momentum in the transverse direction is essential.
While it is mainly handicapped as an analytic closed-form solution is often not accessible, it might be feasible to carry out numerical studies.

On the transverse direction, experimental data from ATLAS~\cite{LHC-atlas-jet-01}, ALICE~\cite{LHC-alice-jet-02}, and CMS~\cite{LHC-cms-jet-01, LHC-cms-jet-02, LHC-cms-jet-03} have shown that the transverse-momentum spectra of hadrons for various collision systems and at various collision energies are very well described by a Tsallis distribution~\cite{ph-tsallis-1, ph-tsallis-2,ph-tsallis-3}.
As a generalization of standard Boltzmann-Gibbs statistics, the Tsallis distribution is derived by maximizing the Tsallis entropy under appropriate constraints.
From the side of data analysis, it properly accounts for the observed slowly decreasing tails obeying the power low.
Besides the phenomenological formula proposed by Hagedorn~\cite{jet-pqcd-17}, a natural recipe is to employ the relativistic hard-scattering model in perturbative QCD.
It was intriguing to observe that the experimental hadron transverse differential cross section appears to differ from what one expects from the intuitive partonic collision viewpoint~\cite{jet-pqcd-10, jet-pqcd-11, jet-pqcd-12}.
To be specific, in the context of quark model, the high-$p_T$ differential cross section in an $2\to 2$ exclusive process can be inferred from a scaling law~\cite{jet-pqcd-10, jet-pqcd-11, jet-pqcd-12}, leading to a different power index from the data.
This fact was subsequently interpreted in terms of the counting rule of the parton-meson degrees of freedom~\cite{jet-pqcd-14, jet-pqcd-15, jet-pqcd-38, jet-pqcd-39}.

The present study, however, concerns an alternative line of thought that does not directly involve the underlying dynamics.
We primarily focus on the phase space distribution regarding the kinematic variables by scrutinizing the partition and constraints of the phase space from a statistical viewpoint.
Specifically, statistics turns out to play an intriguing role in the relevant analyses~\cite{jet-ph-10, jet-ph-12, jet-ph-13, jet-ph-14, jet-ph-15, jet-ph-20, jet-ph-21, jet-ph-22}, where the notion of local partition temperature is a pertinent concept.
For instance, since the system is not homogeneous, it can be argued that the partition temperature fluctuates.
Wilk and Wlodarczyk showed that an additional parameter, $q$, which measures the strength of fluctuations in temperature, readily gives rise to the desired q-exponential distribution~\cite{jet-ph-10}, furnishing a plausible explanation of the data.
Alternatively, if the momentum distribution in an event with fixed multiplicity is featured by a partition function, and the multiplicity fluctuates according to a Gamma distribution, the average momentum distribution obeys also a Tsallis-like form~\cite {jet-ph-12, jet-ph-13}.
Again, in this case, the variance of the multiplicity distribution is associated with the parameters of the Tsallis-type distribution. 
It is thus understood that the apparent validity of such a line of thought indicates that the resulting Tsallis-like distribution is not entirely a manifestation of the underlying dynamics, namely, the QCD. 

It is worth noting that the above two approaches can be viewed consistently under appropriate circumstances.
As pointed out by Wong and Wilk~\cite{jet-pqcd-44}, if one considers the showering and hadronization processes of the parton jets on top of the multiple hard scattering of partons, the experimental observed Tsallis type distributions can be reasonably explained.
On the one hand, by analyzing the experimental results of jet transverse differential cross sections with the relativistic hard-scattering model, it is concluded that the jet production can be approximately described by the relativistic hard-scattering model that appropriately takes into account the effects of multiple scattering and parton thickness.
On the other hand, the showering and hadronization processes further push the power index to a more significant value observed experimentally.
By comparing this picture with the discussions on the statistical side, the latter corresponds to the collinear hadron emission from a given jet.
Employing the phase space integral, the distribution is then estimated at the center-of-mass frame of individual jets, which typically is not of significant relative momentum.
Although the finite size of the underlying phase space implies that the distribution will not have an extensive tail at large momentum, it becomes an exponential function consistent with those of phenomenological models.
Moreover, by convoluting the above two ingredients, it was argued that the resulting distribution is essentially Tsallis type~\cite{jet-ph-12, jet-ph-13}.
In this regard, the present study is qualitative, whose primary goal is to demonstrate the feasibility of effective temperature when the multiplicity is small. 
It is worthwhile to conduct further analysis regarding a quantitative comparison with the observed Tsallis distribution in the data.

\section*{Acknowledgements}

We gratefully acknowledge the financial support from Brazilian agencies 
Funda\c{c}\~ao de Amparo \`a Pesquisa do Estado de S\~ao Paulo (FAPESP), 
Funda\c{c}\~ao de Amparo \`a Pesquisa do Estado do Rio de Janeiro (FAPERJ), 
Conselho Nacional de Desenvolvimento Cient\'{\i}fico e Tecnol\'ogico (CNPq), 
and Coordena\c{c}\~ao de Aperfei\c{c}oamento de Pessoal de N\'ivel Superior (CAPES).
This work is supported by the National Natural Science Foundation of China (NSFC).
A part of this work was developed under the project Institutos Nacionais de Ci\^{e}ncias e Tecnologia - F\'isica Nuclear e Aplica\c{c}\~{o}es (INCT/FNA) Proc. No. 464898/2014-5.
This research is also supported by the Center for Scientific Computing (NCC/GridUNESP) of S\~ao Paulo State University (UNESP).

\appendix

\section{The derivations of the formuale for one-particle distribution function}\label{appA}

In this appendix, we give an account of Eqs.~\eqref{nSim},~\eqref{nCSimZero} and~\eqref{nDSimZero} utilized in the main text.

By plugging Eqs.~\eqref{deltaEpL}, \eqref{deltapT}, and~\eqref{deltaN} into the numerator of Eq.~\eqref{n1Spec} and picking out the terms related to the occupation numbers, we have
\begin{eqnarray}
&& \sum_{\{n_\ell\}} n! \int_0^{2\pi} dv \exp\left[-i\left(n-\sum_{\ell=1}^N n_\ell\right)v\right]\left[\frac{\left(q_1 e^{-(E_1s-p_{L1}t+i \mathbf{p}_{T1}\cdot \mathbf{u}_T)}\right)^{n_1}}{{n_1}!}\right]\cdots\left[\frac{\left(q_Ne^{-(E_Ns-p_{LN}t+i \mathbf{p}_{TN}\cdot \mathbf{u}_T)}\right)^{n_N}}{{n_N}!}\right] \nb\\
&=& \sum_{\{n_\ell\}}n! \int_0^{2\pi} dv e^{-inv} \exp\left[\sum_{\ell=1}^N ivn_\ell \right] \left[\frac{\left(q_1 e^{-(E_1s-p_{L1}t+i \mathbf{p}_{T1}\cdot \mathbf{u}_T)}\right)^{n_1}}{{n_1}!}\right]\cdots\left[\frac{\left(q_Ne^{-(E_Ns-p_{LN}t+i \mathbf{p}_{TN}\cdot \mathbf{u}_T)}\right)^{n_N}}{{n_N}!}\right] \nonumber \\
&=& \sum_{\{n_\ell\}}n! \int_0^{2\pi} dv e^{-inv} \left[\frac{\left(q_1 e^{-(E_1s-p_{L1}t+i \mathbf{p}_{T1}\cdot \mathbf{u}_T)}\right)^{n_1}}{{n_1}!}\right]\cdots\left[\frac{\left(q_Ne^{-(E_Ns-p_{LN}t+i \mathbf{p}_{TN}\cdot \mathbf{u}_T)}\right)^{n_N}}{{n_N}!}\right]\prod_{\ell=1}^N  \left[e^{iv}\right]^{n_\ell}  \nonumber \\
&=& \sum_{\{n_\ell\}}n! \int_0^{2\pi} dv e^{-inv} \prod_{\ell=1}^N  \frac{\left[q_\ell e^{-(E_\ell s-p_{L\ell}t)}e^{iv}\right]^{n_\ell}}{n_\ell!}  \nonumber \\
&=& n! \int_0^{2\pi} dv e^{-inv} \prod_{\ell=1}^N \sum_{n_\ell=1}^\infty \frac{\left[q_\ell e^{-(E_\ell s-p_{L\ell}t+i \mathbf{p}_{T\ell}\cdot \mathbf{u}_T)} e^{iv}\right]^{n_\ell}}{n_\ell!} \nb\\
&=&  n! \int_0^{2\pi} dv e^{-inv}\prod_{\ell=1}^N\exp\left[q_\ell e^{-(E_\ell s-p_{L\ell}t+i \mathbf{p}_{T\ell}\cdot \mathbf{u}_T)} e^{iv}\right], \label{pluggingAll}
\end{eqnarray}
where the remaining integral in $v$ will still turn out to be manageable.
A similar derivation can be carried out for the denominator of Eq.~\eqref{n1Spec}.

To proceed, we note that the part involving the product $\prod_{\ell=1}^N$ between phase space intervals can be wrapped up as follows
\begin{eqnarray}
\prod_{\ell=1}^N\exp\left[q_\ell e^{-(E_\ell s-p_{L\ell}t+i \mathbf{p}_{T\ell}\cdot \mathbf{u}_T)} e^{iv}\right]
= \exp\left[\sum_{\ell=1}^N q_\ell e^{-(E_\ell s-p_{L\ell}t+i \mathbf{p}_{T\ell}\cdot \mathbf{u}_T)} e^{iv}\right],
\end{eqnarray}
and therefore
\begin{eqnarray}
\lim\limits_{N\to\infty}\prod_{\ell=1}^N\exp\left[q_\ell e^{-(E_\ell s-p_{L\ell}t+i \mathbf{p}_{T\ell}\cdot \mathbf{u}_T)} e^{iv}\right]
= \exp\left[F(s,t,\mathbf{u}_T) e^{iv}\right],
\end{eqnarray}
where
\begin{eqnarray}
F(s,t,\mathbf{u}_T) 
\equiv \lim\limits_{N\to\infty}\sum_{\ell=1}^N q_\ell e^{-(E_\ell s-p_{L\ell}t+i \mathbf{p}_{T\ell}\cdot \mathbf{u}_T)}
=dyd\mathbf{p}_Tf(y,\mathbf{p}_T) \exp\left[-\sqrt{\mathbf{p}_T^2+m^2}(s \cosh y- t \sinh y)+i\mathbf{p}_T \cdot \mathbf{u}_T\right]. \label{Fdef} \nb\\
\end{eqnarray}

The resulting contribution that involves the integral in $v$ can be dealt with analytically. 
For the numerator, it possesses the following form
\begin{eqnarray}
&&\int_0^{2\pi} dv \exp[-inv] \exp\left[F(s,t,\mathbf{u}_T) e^{iv}\right]\nb\\
&=&  \int_0^{2\pi} dv \exp[-inv] \sum_{l=0}^\infty \frac{F(s,t,\mathbf{u}_T)^l e^{ilv}}{l!} \nb\\
&=&  \sum_{l=0}^\infty \frac{1}{l!} \int_0^{2\pi} dv \exp[-i(n-l)v]F(s,t,\mathbf{u}_T)^l  \nb\\
&=&  \sum_{l=0}^\infty \frac{F(s,t,\mathbf{u}_T)^l}{l!} \delta_{n,l}  \nb\\
&=&  \frac{F(s,t,\mathbf{u}_T)^n}{n!} .
\end{eqnarray}

By putting the above pieces together, one simplifies Eq.~\eqref{n1Spec} into the following forms~\cite{jet-ph-05}
\begin{eqnarray}
A &=& \frac{-n}{(2\pi)^4}f(y,\mathbf{p}_T)dyd\mathbf{p}_T\int_{\epsilon_0-i\infty}^{\epsilon_0+i\infty} ds\int_{\epsilon_1-i\infty}^{\epsilon_1+i\infty}dt\int du_T \left[F(s,t,u_T)\right]^{n-1} \nonumber \\
&\times& \exp\left[(W-\sqrt{\mathbf{p}_T^2+m^2}\cosh y)s-(P_L-\sqrt{\mathbf{p}_T^2+m^2}\sinh y)t-i(\mathbf{P}_T-\mathbf{p}_T)\cdot u_T\right],
\end{eqnarray}
and
\begin{eqnarray}
B=-\frac{1}{(2\pi)^4}\int_{\epsilon_0-i\infty}^{\epsilon_0+i\infty}ds\int_{\epsilon_1-i\infty}^{\epsilon_1+i\infty}dt\int du_T \left[F(s,t,u_T)\right]^{n} \exp\left[Ws-P_Lt-i\mathbf{P}_T\cdot u_T\right] .
\end{eqnarray}

As elaborated further below~\cite{jet-ph-12, jet-ph-13}, if these hadrons are emitted from a jet, it is a reasonable approximation to assume that $\langle p_T\rangle$ is not significant when compared to $\langle p_L\rangle$ as the hadrons are mostly aligned. 
This is in accordance with the scenario of di-jet when the momentum component in the perpendicular direction of the jet is insignificant.
In this regard, one relaxes the conservation of transverse momentum by assuming that $F(s,t,\mathbf{u}_T)$ does not depend on $\mathbf{u}_T$.
We therefore replace $F(s,t,\mathbf{u}_T)$ by $F(s,t)$ in the above expressions.
By noticing that the integrations in $\mathbf{u}_T$ become irrelevant and cancel out in the ratio, we find
\begin{eqnarray}
\left.\frac{d^3\bar{n}}{dyd\mathbf{p}_T}\right|_{n,W,P}
\equiv\langle n_k \rangle 
\simeq nf(y,\mathbf{p}_T)dyd\mathbf{p}_T\frac{C}{D} ,\label{nSimApp}
\end{eqnarray}
where
\begin{eqnarray}
C = \int_{\epsilon_0-i\infty}^{\epsilon_0+i\infty}ds\int_{\epsilon_1-i\infty}^{\epsilon_1+i\infty}dt\left[F(s,t)\right]^{n-1} 
\times \exp\left[(W-\sqrt{\mathbf{p}_T^2+m^2}\cosh y)s-(P_L-\sqrt{\mathbf{p}_T^2+m^2}\sinh y)t\right] ,\label{nCSim}
\end{eqnarray}
and
\begin{eqnarray}
D=\int_{\epsilon_0-i\infty}^{\epsilon_0+i\infty}ds\int_{\epsilon_1-i\infty}^{\epsilon_1+i\infty}dt\left[F(s,t)\right]^{n} \exp\left[Ws-P_Lt\right] .\label{nDSim}
\end{eqnarray}

By assuming $m=0$ in Eqs.~\eqref{nCSim} and~\eqref{nDSim}, one readily obtains Eqs.~\eqref{nCSimZero} and~\eqref{nDSimZero} given in the main text.

\bibliographystyle{h-physrev}
\bibliography{references_qian}

\end{document}